\documentclass[pdftex,aps,twocolumn,prl,showpacs,color,psfig]{revtex4}
\usepackage{amsmath}
\usepackage{color}
\usepackage{amsfonts}
\usepackage[pdftex]{graphicx}
\begin{document}

\title{Bistable defect structures in blue phase devices}

\author{A. Tiribocchi$^1$, G. Gonnella$^1$, 
D. Marenduzzo$^2$, E. Orlandini$^3$, F. Salvadore$^4$}

\affiliation{
$^1$ Dipartimento di Fisica {\rm and} Sez.~INFN di Bari, Universit\`a di Bari, 
 70126 Bari Italy \\
$^2$ SUPA, School of Physics, University of Edinburgh, 
Edinburgh EH9 3JZ, Scotland\\
$^3$ Dipartimento di  Fisica {\rm and} Sez.~INFN di Padova, Universit\`a di Padova, 35131 
Padova Italy\\
$^4$ Caspur - via dei Tizii 6/b, 00185 Roma Italy}

%\author{A. Tiribocchi} 
%\affiliation{Dipartimento di Fisica {\rm and} Sez.~INFN di Bari, Universit\`a di Bari, 70126 Bari Italy}

%\author{G. Gonnella} 
%\affiliation{Dipartimento di Fisica {\rm and} Sez.~INFN di Bari, Universit\`a di Bari, 70126 Bari Italy}

%\author{D. Marenduzzo}
%\affiliation{SUPA, School of Physics, University of Edinburgh, 
%Edinburgh EH9 3JZ  UK}

%\author{E. Orlandini}
%\affiliation{Dipartimento di  Fisica {\rm and} Sez.~INFN di Padova, Universit\`a di Padova, 35131 Padova Italy}

\begin{abstract}
Blue phases (BPs) are liquid crystals made up by networks of defects, or disclination lines. While existing phase diagrams show a striking variety of competing metastable topologies for these networks, very little is known as to how to kinetically reach a target structure, or how to switch from one to the other, which is of paramount importance for devices. We theoretically identify two confined blue phase I systems in which by applying an appropriate series of electric field it is possible to select one of two bistable defect patterns. Our results may be used to realise new generation and fast switching energy-saving bistable devices in ultrathin surface treated BPI wafers.

\if{
Blue phases (BPs) are liquid crystalline phases made up by networks of defects, or disclination lines, which
form the basis of the new generation of fast switching display devices. Here we show that, by
exploiting the presence of several metastable disclination networks with competing free energy, blue phases
may be used to build energy-saving bistable devices, which can retain either of two states in the absence of an
applied electric field. We identify two such devices, both based on blue phase I: the first exploits surface memory
effect, while the second relies on the interplay between homeotropic anchoring and confinement, and may be
built in practice by using ultrathin surface treated BP wafers.}\fi
\end{abstract}

\maketitle

Liquid crystals are materials with spontaneously broken symmetry: as such they support topological defects of various kinds and may be used as testing grounds for theories on many areas of physics, e.g. the dynamics of cosmic strings and of vortex lines in superfluids and superconductors \cite{Other}.
Blue phases (BPs) are a spectacular example of soft matter in which a network of such defects self assembles as a fully
periodic cubic three-dimensional structure whose local director has a double-twist configuration~\cite{mermin}. The defects are arranged in a regular fashion, with periodicity at length scales comparable with the wavelenght of light. This is the reason why BPs
exhibit selective Bragg reflection in the range of visible light and can have 
interesting applications in fast light modulators~\cite{dim89} 
and tunable photonic crystals~\cite{photonic}.
BPs were initially observed to be stable only in a narrow range of temperature $\sim 1K$, but
recently new compounds, stable in wider interval over $\sim 60K$ (including
room temperature), have been developed~\cite{kiku,coles}. These
studies can be considered with good reason as seeds for the recent
fabrication of the first blue phase display device with fast switching
times~\cite{sams}. On the other hand BPs have attracted broader interest among physicists
because double-twist cylinders represent a striking example
of the so-called skyrmions, topological excitations encountered in
nuclear physics~\cite{nucl_brown}, in spinor Bose-Einstein condensates~\cite{usama}, 
and in ferromagnetic materials, where they have been %also been 
observed~\cite{science}.

Theory and experiments have mapped out a number of phase diagrams reporting a remarkably varied range of competing metastable disclination networks \cite{mermin,dupuis}. However, there are only very few theoretical studies to date regarding the dynamics of the formation of BPs, and virtually none addresses how to switch from one state to another reversibly in a controlled way. Here we provide the first theoretical study to achieve this goal, and give two specific examples of a bistable BPI system, where an electric field along the appropriate direction can select one of two metastable arrangement of defects. Besides being of interest {\it per se}, as an intriguing example of controlled defect dynamics, our discovery suggests specific novel geometries which may be used to realise bistable BP based devices. 
Devices like the ones we have found are important technologically as each of the bistable states is metastable and is retained after the field is switched off: this fact eliminates the need of having a constant electric field to keep the system  in the \lq\lq on\rq\rq state, hence sharply reducing energy consumptions. 
Some examples of such systems are the zenithal bistable device (ZBD)~\cite{zenit} and surface-stabilised cholesteric texture (SSCT) devices~\cite{kent}. \if{In both these, one of competing metastable states is defect free, while the other contains disclinations.}\fi In the ZBD case bistability relies on flexoelectricity and requires a periodic grating of the surfaces, whereas in the SSTC case it exploits the existence of focal conic defects, which compete against the defect-free helical texture when strong anchoring is used.  Several other nematic multistable devices also exist (see e.g.~\cite{nem_dev}). Our case is different as the switching occurs between two disclination networks. BP devices such as the one we suggest should also share the faster switching times made possible by BP cells. At the same time, simulations are key to clarify the dynamics of the defect reorganisation underpinning bistability~\cite{zenit,nem_dev}. 

In this work we propose two novel BPI (blue phase with $O_8^{-}$ symmetry) geometries, for which it is possible to design a simple switching on-off schedule that leads to a bistable defect network.
In both cases, bistability is present for a sizeable range of parameters (quantified below) and is characterized by the presence of two states, which are both metastable when the field is off and have a fully 3D defect structure. 
The first device we propose relies on surface memory effects which retain the BPI structure on the boundaries, and the two competing states are the standard bulk BPI network and another arrangement, which is only slightly higher in free energy. The second device exploits the additional frustration driven by strong homeotropic anchoring in a confined thin sample, which conflicts with the bulk ordering. We find that this leads to the appearance of multiple defect states, similar to the ones observed in~\cite{fukuda2}(a), which, importantly, can be uniquely selected by the application of an electric field along an appropriate direction.
Our devices are entirely different from the BP bistable device described in~\cite{wang}. 
Firstly, in that device the dielectric anisotropy is negative, while in our case
it is positive. Secondly, the switching in~\cite{wang} is between two field-induced metastable states,
which are selected by different values of an applied voltage, whereas in our case the electric fields is
switched on and off along different directions.
%Finally we quantify the effect of backflow for this device showing that it decrease the relaxation
%times in the evolution towards stationary states.

The physics of a BP device may be described by a Landau-de Gennes free energy 
written in terms of the tensor order parameter $Q_{\alpha\beta}$~\cite{mermin}.
This comprises a bulk term
%\begin{equation}
$f_{b}  =  \frac {A_0}{2} (1 - \frac {\gamma} {3}) Q_{\alpha \beta}^2 - 
          \frac {A_0 \gamma}{3} Q_{\alpha \beta}Q_{\beta
          \gamma}Q_{\gamma \alpha} %\\ \nonumber
         +   \frac {A_0 \gamma}{4} (Q_{\alpha \beta}^2)^2$,
%\label{eqBulkFree}
%\end{equation}
and a distortion term \cite{mermin}
%\begin{equation}
$f_{d} = \frac{K}{2} (\partial_\beta Q_{\alpha \beta})^2
+ \frac{K}{2} (\epsilon_{\alpha \zeta\delta }
\partial_{\zeta}Q_{\delta\beta} + 2q_0Q_{\alpha \beta})^2$, %\\
%\end{equation}
where $K$ is the elastic constant and the pitch of the cholesteric liquid crystal is
given by $p\equiv 2\pi/q_0$. $A_0$ is a constant, $\gamma$ controls the 
magnitude of the ordering and $\epsilon_{\alpha \zeta\delta}$ is the Levi-Civita
antisymmetric third-rank tensor.
% $A_0$ is a constant and $\gamma$ controls the 
%magnitude of the ordering. 
The constants $A_0$, $K$ and $\gamma$ are chosen in order to be in the appropriate
region of the phase diagram~\cite{dupuis}. We have set $A_0=0.0034$, $K=0.005$, $\gamma=3.775$
for which BPI is stable.
%, and $A_0=0.00212$, $K=0.005$, $\gamma=2.9$ for BPII. 
This choice may be mapped to a rotational viscosity equal to 1 Poise, (Frank)
elastic constants equal to $\sim 30$ pN and a blue phase periodicity equal to 300 nm.
%These may correspond, for instance, to a blue phase with lattice constant equal to 400 nm, 
%with a rotational viscosity equal to 1 Poise and (Frank) elastic constants equal to $\sim 10$ pN 
%and $\sim 16$ pN for BPI and BPII respectively. 
%Accordingly, 
Moreover one space and time unit correspond to 0.0125 $\mu$m and 0.0013 $\mu$s. % for BPI, and to 0.0125 $\mu$m and 
%0.0017$\mu$s for BPII.
The contribution of the boundaries is encoded in a surface free-energy term, given by $f_s=1/2W_0(Q_{\alpha\beta}-
Q_{\alpha\beta}^0)^2$, where $W_0$ (larger than $ 0.01~N/m$ in our simulations) 
controls the anchoring strength and $Q^0_{\alpha\beta}$ is the order parameter at the surfaces.
The interaction with an external electric field ${\bf E}$ is accounted for by an additional term,
$-\frac{\varepsilon_a}{12\pi}E_{\alpha} Q_{\alpha \beta} E_{\beta}$, where
$\varepsilon_a$ (here taken $>$0) is the dielectric anisotropy of the material.
The electric field may be quantified via the dimensionless number $e^2=27 \varepsilon_a E_{\alpha}^2/(32\pi A_0\gamma)$.
By assuming %$\frac{27}{2A_0\gamma}\sim 10^{-5}J^{-1}m^3$~\cite{mermin}
a dielectric anisotropy of $\sim 100$~\cite{rao}, one finds that an electric field of $1-10~V/\mu m$ corresponds to
$e^2\sim 0.003 - 0.3$.
%whereas an electric field of $10 V/\mu m$ corresponds to $e^2\sim 0.2$.}
The simulation domains along the three coordinate axes are $L_x$, $L_y$ and $L_z$ respectively,
with periodic boundaries along $\hat{x}$ and $\hat{y}$. All our results are performed
on lattices of dimensions $32\times 32\times 32$.
The equation of motion for {\bf Q} is  $D_t{\bf Q}= \Gamma {\bf H}$, where $\Gamma$ is a
collective rotational diffusion constant and $D_t$ is the material derivative for rod-like molecules~\cite{beris}. 
The molecular field $\mathbf{H}$ is the thermodynamic force that
drives the system towards the equilibrium~\cite{beris}. The velocity field obeys a Navier-Stokes
equation with a stress tensor generalised to describe LC hydrodynamics~\cite{beris}.
The interplay between the velocity field and the order parameter is referred to as
backflow. To solve the equations of motion, we used a hybrid lattice Boltzmann algorithm, as 
in~\cite{nem_dev}. 

The usual zero-field defect structure in the BPI cell is shown in Fig.~1a. 
We first focus on the case in which the director field is fixed at the top ($z=L$, being $L$ the sample size) 
and bottom ($z=0$) boundaries of the cubic cell to its stable structure in absence of a field.
This may be achieved by pinning due to impurities or through surface memory effects~\cite{helfr2}.
\begin{figure}
\centerline{\includegraphics[width=8.5cm]{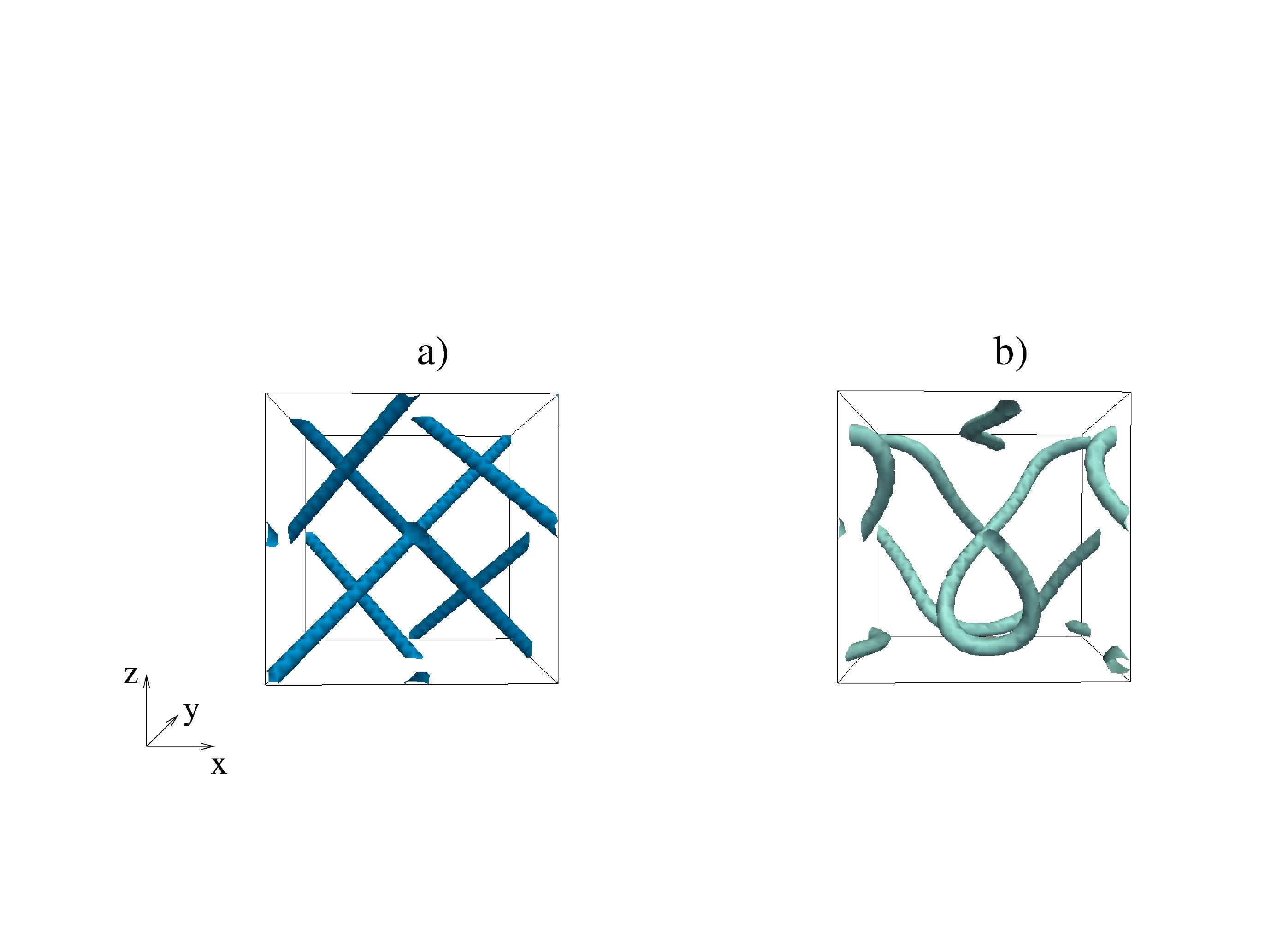}}
\caption{Equilibrium disclination networks for BPI when a) the structure is
strongly anchored at both walls, and when b) there is homeotropic anchoring
at both walls.}
\label{fig1}
\end{figure}
Fig.~2a-d shows the evolution of the defect network %of the BPI device
in response to an electric field switched on and off first along $\hat{x}$, 
then along $\hat{y}$. During the cycle the defect dynamics is highly non-trivial and 
is characterized by a complex reorganization driven by the electric field, and depending on its 
direction. In particular, after the application of the field along $\hat{x}$, 
the defects twist and bend and eventually recombine,
annihilating in the bulk and forming defect arcs pinned at the walls ($t_1-t_4$). The voltage 
($\Delta V_y=E_yL_y$)
is chosen strong enough to break the defect structure in the bulk. When the field is switched off,
some disclinations migrate back to the centre of the cell, giving rise to connected boundary arcs 
that span the entire device ($t_5-t_8$). The system gets stuck into a stationary metastable state ($t_8$),
whose free energy is different from the one of the zero-field state (see inset of Fig.~\ref{fig2}, top panel).
Due to the anisotropy of the BPs, it is now interesting to ask what is the defect dynamics 
if the electric field acts on a different direction.
Starting from the configuration at time $t_8$ we have switched on an electric field along the $\hat{y}$ direction,
with voltage equal to the one chosen for the field in $\hat{x}$ direction. The defects now start to deform, bending
in the direction of the field ($t_9$), and then move towards the bulk where they met and form unstable branching
points that annihilate. This second field-induced state ($t_{12}$) is again (as the one at time $t_4$) characterized
by boundary arcs. By switching off the field, the system reorganizes as the boundaries force the disclinations
towards the bulk: surprisingly, the dynamics now leads back to the starting bulk BPI (Fig.~1a). 
Bistability is affected by the voltages used in the switching. We have simulated the same device
for $e^2\sim 0.2, 0.3$ and found that it is bistabile in both cases. On the other hand, 
bistability is lost for $e^2<0.2$ and the switching dynamics is now characterized by a reversible cycle 
in which a zero-field state (different from the bulk BPI structure, Fig~1a), 
obtained after the application of the electric field along $\hat{x}$, 
is recovered even after switching along $\hat{y}$. Note also that the results in Fig. 2 refer to
BP devices with $\varepsilon_a>0$: intriguingly, in the case of negative dielectric constant the device
can never get back to the bulk BPI state. This may be due to the fact that in the presence of a field the 
director can form a helix along the field, and this is not different enough in free energy from the
BP state to drive a reorganisation of the unit cell once the field is removed.
\if{In this case the device can be considered as switchable, similarly
to what has been observed in the Ref.~\cite{soft} with the field perpendicular to the walls.}\fi
To check the robustness of our results, we have also simulated the device with
different values of $A_0$ and $\gamma$ (e.g. $A_0=0.0012$ and $\gamma=5.085$), again 
leading to a stable BPI -- even in this different region of phase space we find that the device is bistable.
Finally, in Fig.~2, top panel, we also  compare the free energies as a function of time both in the presence and
absence of hydrodynamics. The effect of backflow is to slightly speed up all relaxations, 
without deep consequences on the dynamic evolution.

\begin{figure}
\centerline {\includegraphics[width=17cm]{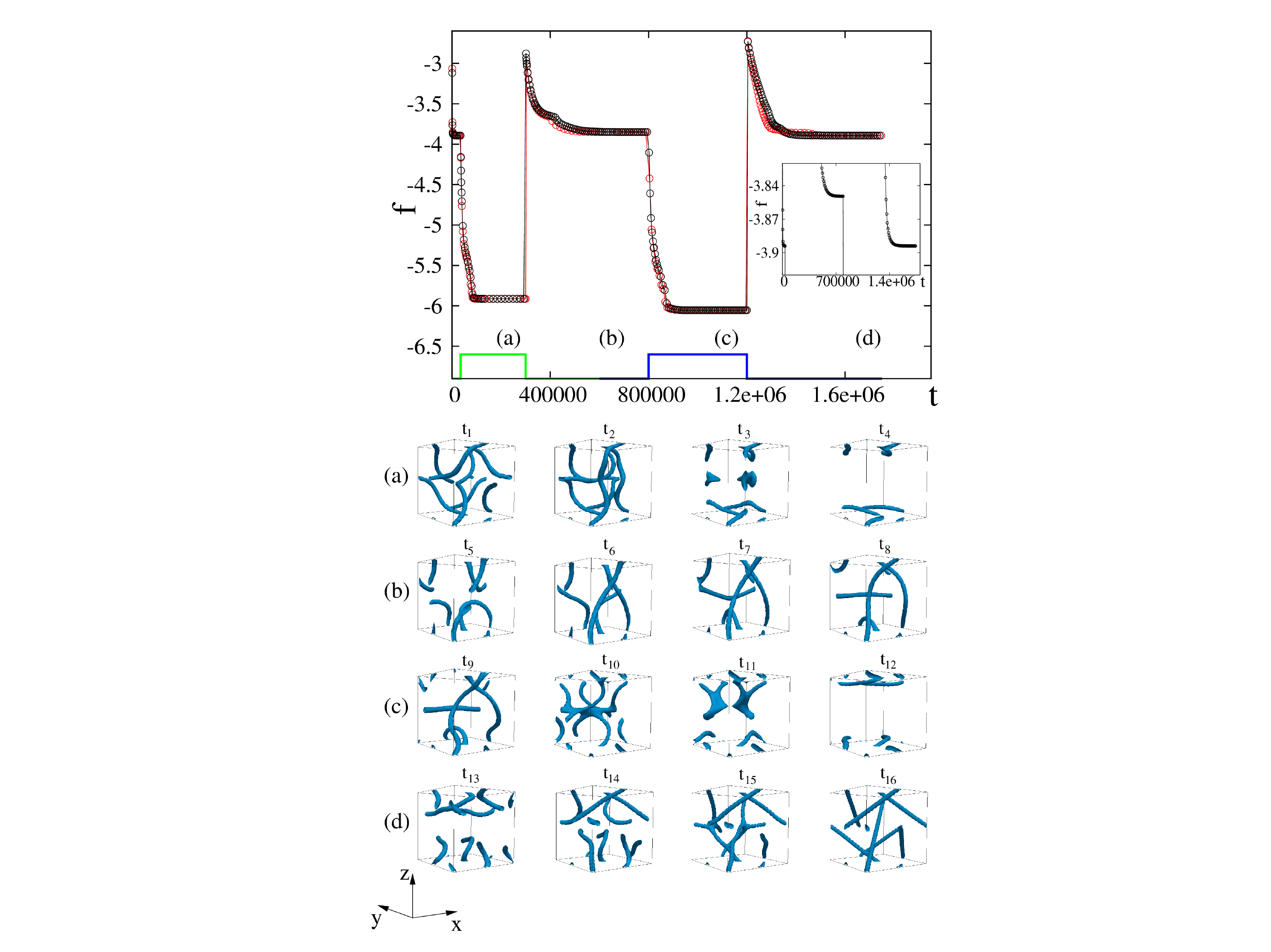}}
\caption{Free energy evolution (top panel) of the BPI device with FBCs, with (red) and without (black) backflow.
Parameters were $A_0=0.0034$, $K=0.005$, $\gamma=3.775$, $e^2\sim 0.2$ (corresponding to a field of $\sim 8~V/\mu m$). 
The evolution of the defects under an applied field is reported in the bottom panel. 
The step function in the top panel is $\ne 0$ when the field is on.
From $t=0$ up to $t=35000$ the system relaxes to the stable BPI configuration in Fig.~1a.
An electric field along $\hat{x}$ is switched on at $t=35\times 10^3$ and 
the equilibrium field-induced state is reached at $t_4=30\times 10^4$.
The corresponding free-energy is shown in (a), top panel.
Then the field is switched off and the system 
relaxes to a new metastable state ($t_8=80\times 10^4$), ((b) in the bottom panel)
characterized by a new free energy, slightly higher than the bulk BPI value.
An electric field along $\hat{y}$ is then switched on and off at $t_{12}=120\times 10^4$
((c) and (d) in the top panel). Strikingly, 
the relaxation process brings the system back to the equilibrium state without field.} 
\label{fig2}
\end{figure}

\begin{figure}
\centerline {\includegraphics[width=12cm]{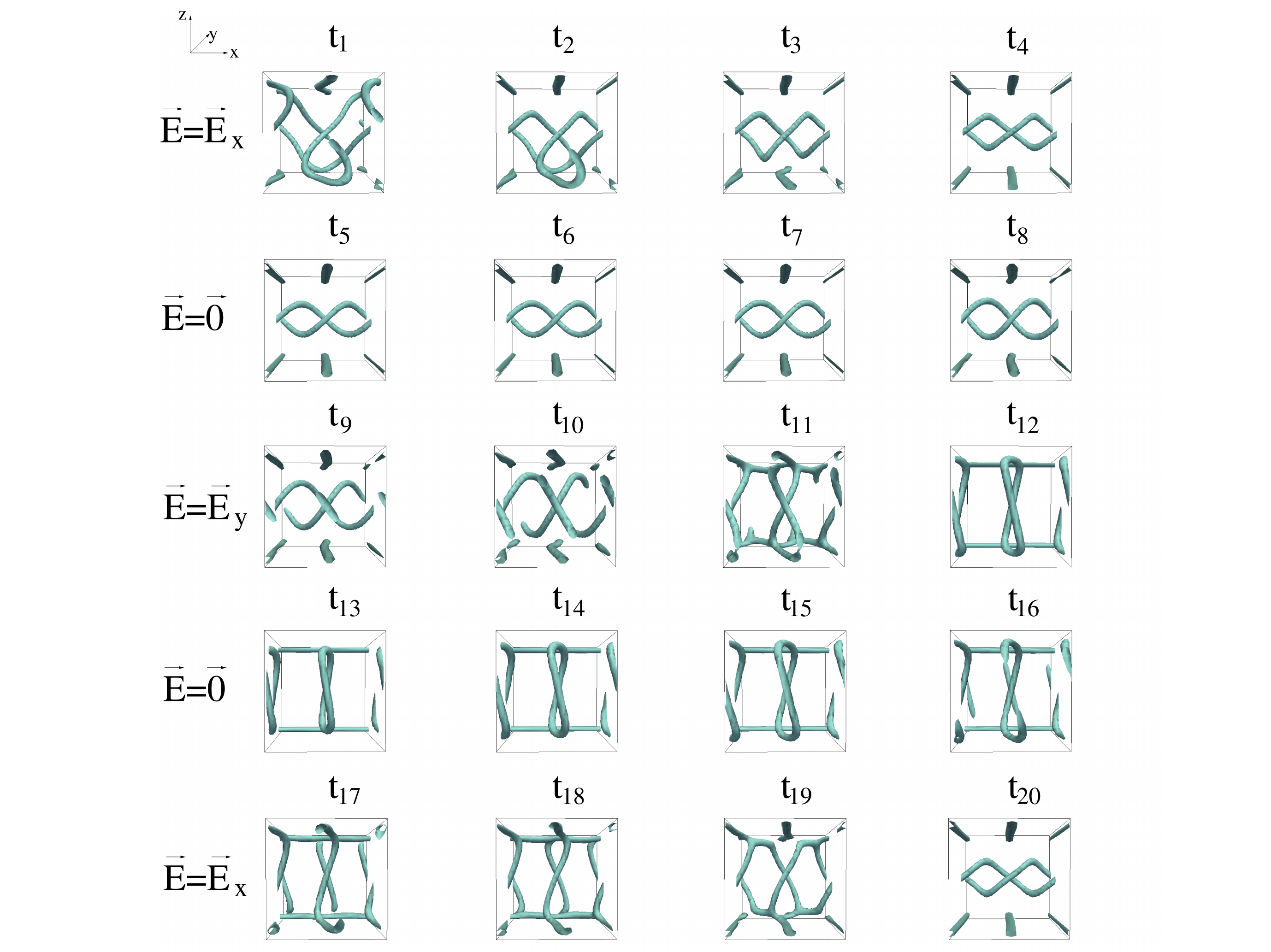}}
\caption{
Defect evolution with homeotropic anchoring under an applied electric field.
Parameters are the same as in Fig.~\ref{fig2}.
The electric field ($\sim 4.5~V/\mu m$, i.e. with $e^2\sim 0.06$) is switched on along $\hat{x}$
at time $t=10\times 10^4$ (Fig.~\ref{fig1}b for the steady state)
and the evolution dynamics is reported in the first row. A steady state double-helix defect is attained  
at time $t_4=98.7\times 10^4$.
%$t_1=61\times 10^4$, $t_2=79.7\times 10^4$, $t_3=81.7\times 10^4$, $t_4=98.7\times 10^4$.
When the field is switched off the defect structure is almost 
unaltered (steady state reached at time $t_8=117.7\times 10^4$).
%$t_5=99.7\times 10^4$, $t_6=109.7\times 10^4$, $t_7=110.7\times 10^4$, $t_8=117.7\times 10^4$.
Then field ($\sim 4~V/\mu m$, i.e. with $e^2\sim 0.05$) is then switched on along $\hat{y}$. The double-helix disclination
transforms into a steady state with twisted ring defects ($t_{12}=155.6\times 10^4$)
stable after switching off ($t_{16}=194.2\times 10^4$).
Bistability requires a route back to the double helix state: this is provided by a further application
of a field along $\hat{x}$ (steady state at $t_{20}=223.4\times 10^4$).
}\label{fig3}
\end{figure}

We now turn to the case of different boundary conditions. Fixed boundary conditions (FBCs), as the ones we have just employed, have often been used in cholesterics and BPs~\cite{helfr2}. 
%While the dynamics leading to bistability is 
%theoretically intriguing, these boundary conditions are difficult 
%to be accurately realised in practice, as they rely on surface impurities. 
However in practice it is more feasible to control
 the alignment of the director field at the boundaries by chemical treatment or rubbing. The former technique typically is used to enforce homeotropic (or perpendicular) anchoring of the director at the wall, the latter yields homogeneous (or parallel) anchoring. Therefore it is interesting to know
how these boundaries affect the switching dynamics of a BP device, and, importantly, whether
they can still lead to a bistable behaviour. In what follows we focus on a BPI device with
{\em homeotropic} anchoring at both walls. 
To the best of our knowledge there is only
one study about the equilibrium disclination configuration in confined BPI cell with homeotropic anchoring~\cite{fukuda2}(a)
%and another with homogeneous anchoring~\cite{fukuda3},
while nothing is known about electric field effects in these thin cells.
As we shall see, by using the same electric field dynamical schedule
adopted for the cell with FBCs, it is still possible to realise a switchable bistable device, albeit with some key
conceptual differences with respect to the device in Fig. 1a.
The first one is that homeotropic anchoring affects the zero-field equilibrium defect 
structure (Fig.~\ref{fig1} b)), now characterized by bent disclinations depinned from the walls -- 
the original topology of BPI being preserved.
The second important difference  is that the change of the boundary conditions affects the dynamic evolution 
under an electric field: the minimum value required to destabilise the initial topology is now $e^2\sim 0.05$.
This decrease in the threshold is because the wall now pins the zero-field structure more weakly than in the
case of Fig. 1.
In spite of these differences, there is a range in which the device shows bistability, and the
switching dynamics of the defect network is reported in Fig.~\ref{fig3}. When the electric field is switched on along
the $\hat{x}$ direction ($t_1-t_4$), the defects start to deform, bend and form a spectacular double-helix pattern in the bulk, accompanied by columnar defects at the wall.
This structure is induced and stabilised by the electric field, yet it only suffers minor rearrangements as the field is
switched off. This double-helical state has
already been observed in Ref.~\cite{fukuda2}(a) in the {\em absence}
of a field -- it could be selected for instance by tuning the device thickness.
When we switch on the field again, now along the $\hat{y}$ direction (see snapshots $t_9-t_{16}$), 
the bent double-helix moves towards the walls ($t_9-t_{10}$) where it joins up with the the columnar disclinations,
to form a transient interconnected structure ($t_{11}$) which later on relaxes to an array of 
twisted ring defects in the bulk, accompanied again by straigth disclinations, this time at right angles with the one
associated with the double helical state ($t_{12}$). 
Once more the field-induced rings do not change much upon switching off ($t_{12}-t_{16}$). 
We note that a state similar to these ring defects has recently been observed in 
Ref.~\cite{fukuda2}(b), however in a system with {\em homogeneous} anchoring at the walls.
Our simulations therefore show that the number of possible defect networks in these thin BP cells are
even more than previously thought.
In order to behave as a switchable bistable device, our system still needs to be able to reconstruct the double-helical
state starting from the ring defect structure. This is indeed what happens if the field is switched on and off 
along the $\hat{x}$ direction (as in the first cycle). The ring defects initially enlarge
and touch the walls where disclination lines reform ($t_{17}-t_{18}$). 
Thereafter the double-helix pattern reconstructs in the bulk ($t_{19}$) and later on stabilises.
Thus our thin BP sample with homeotropic anchoring behaves as a bistable device, and the two switchable states
are the double helical and twisted ring structures -- unlike the surface memory device none of these equals
the zero-field stable state.

In summary, we have investigated the hydrodynamics of a cubic BPI cell
and proposed two systems, differing in the boundary conditions, which can switch
between two bistable disclination structures under the action of an applied field.
In the first one we have fixed boundary conditions at both walls, while in the second one we considered
homeotropic anchoring. 
\if{In both cases the switching dynamics is strongly affected by the magnitude and by the direction of an applied electric field, whose 
schedule is fundamental to observe bistability.}\fi
In the FBCs cell, the two switchable states are the zero-field BPI and another metastable defect network.
The device with homeotropic anchoring, on the other hand, switches between two states which are metastable in the
absence of a field. In one of these states the defects form a double helix, whereas in the other one they
arrange as an array of twisted rings -- these configurations are alternatively observed by varying the direction
and the magnitude of the electric field. The bistable behaviour of this device is promising 
for technological applications to energy-saving devices because, in principle, the required anchoring could 
easily be achieved. 
The physical reason behind the viability of BP bistable devices may be the glassy free energy profile which admits a number of competing metastable minima, each of which is potentially suited for their design, provided a suitable kinetic route to achieve reversible switching is discovered. In this respect, one may expect that even more BP bistable systems may be found in the future.

\end{document}